# Ultra-broadband Optical Switching Plasmons Waveguide in Ge Nanowires

Xinghui Liu[1,2,†], Kaili Chang[3,†], Jiarong Guo[1,5,†], Mengfei Xue[4,†], Ran Zhou[3], Ke Chen[3,*], Jianing Chen[1,5,6,*]

1. Beijing National Laboratory for Condensed Matter Physics, Institute of Physics, Chinese Academy of Sciences, Beijing 100190, China.
2. State Key Laboratory of Quantum Optics and Quantum Optics Devices, Institute of Laser Spectroscopy, Collaborative Innovation Center of Extreme Optics, Shanxi University, Taiyuan, Shanxi 030006, China.
3. Center for the Physics of Low-Dimensional Materials, School of Future Technology, School of Physics and Electronics, Henan University, Kaifeng, 475004, China.
4. Suzhou Laboratory, Suzhou 215100, China.
5. University of Chinese Academy of Sciences, Beijing 100049, China.
6. Songshan Lake Materials Laboratory, Dongguan, Guangdong 523808, China.
†These authors contributed equally to this work.
*E-mails: kchen@henu.edu.cn, jnchen@iphy.ac.cn.

**Abstract**

Plasmonic devices, with their ultra-high integration density and data-carrying capacity comparable to optical devices, are currently a hot topic in the field of nanophotonic devices. Photodetectors, non-volatile memories, and ultra-compact lasers based on plasmons in low-dimensional materials are emerging at a rapid pace. However, the narrow optical response band and limited of convenient tunable methods currently available have hindered the development of these plasmonic materials. Here, we report a ultrabroadband non-equilibrium plasmonic responses based on Ge nanowires tuned by optical method. We tracked the blue shift of the plasmonic response of Ge nanowires due to photo-induced carriers over an ultra-broad spectral range of 800 – 2000 cm$^{-1}$. For the first time, we have achieved the imaging of propagating surface plasmon polaritons (SPPs) in semiconductor nanowires, which were tuned by photo-induced carriers. The ultrafast and ultrabroadband response of semiconductor nanowire plasmons is of great significance for future ultrafast all-optical devices.

**Introduction**

Plasmons are collective oscillations of electrons and the electromagnetic field in a material, which arise from the coupling between the oscillating electrons and photons.[1–3] Plasmonic devices possess advantages of both photonic and electronic devices, thanks to their intrinsic hybrid property and the highly compressed wavelength they offer.[4] The highly compressed wavelength of plasmons, on the order of nanometers, facilitates the high integration of plasmonic devices.[5] Additionally, plasmons operate at optical frequencies, allowing for data carrying capacity comparable to that of photonic devices.[6,7] While noble metals are the most extensively studied plasmonic platforms, their high initial electron densities and lossy nature can limit the electronic tunability and broader applications of metallic plasmons.[8–10] The emergence of Van der Waals materials in the past decade has inspired research and applications in plasmonic with multi-field tunability.[2,11,12] For instance, researchers have demonstrated the efficient tuning of the Dirac plasmon in graphene through various techniques, including electrical gating[13,14], chemical doping[15,16], optical excitation[17,18], and multiple layer coupling[19,20]. Besides versatile Van der Waals materials, the plasmonic behaviors of semiconductors can also be readily manipulated by tuning their carrier properties.[21–24] Given the well-developed semiconductor industry, semiconductors have broader and more direct applications in on-chip plasmonic optoelectronic devices, including detectors[25,26], memories[27,28] and lasers[29,30]. Due to their high carrier mobility, the direct narrow bandgap III-V semiconductors InAs and InSb have become excellent mid-infrared plasmonic platforms and are widely used in infrared-terahertz detection.[24,31,32] The previous investigation on InAs and InSb nanowires have demonstrated its active optical tunability and low loss[21,24]. However, the low upper limit of optical carrier incubation under the hv = 0.8 eV laser excitation prevented the tunable range of their plasmon response and the visualization of surface plasmon polaritons (SPPs), limits their application in ultrafast optoelectronic devices.

Ge has the highest known carrier mobility of p-type doped semiconductors and has good compatibility with III-V semiconductors, and has attracted great attention in non-silicon-based semiconductor devices in recent years[33,34]. Ge is an indirect bandgap semiconductor, but its direct bandgap is only slightly higher than the indirect bandgap.[35] This gives us the opportunity to study the non-equilibrium plasmons of Ge nanowires by optically tuning the carrier concentration of Ge. In this paper, the excitation energy of 1.6 eV excites the multi-valence band electron transitions of Ge nanowires containing heavy holes (HH), light holes (LH), and split-off holes (SOH), which greatly improves the upper limit of photo-induced carriers. We traced the blue shift of the Ge nanowire plasmonic response due to photo-induced carriers over a span of more than 1000 cm$^{-1}$(800 – 2000 cm$^{-1}$). By combining near-field broadband imaging and hyperspectral line scanning, we have, for the first time, achieved imaging of the propagating SPPs of semiconductor nanowires tuned by photo-induced carriers. The ultrafast and ultrabroadband response of semiconductor nanowire plasmons is of great significance for future ultrafast optical logic devices. The ultrafast and ultrabroadband response of semiconductor nanowire plasmons is of great significance for the development of future ultrafast optical logic devices.

**Results**

We employed an ultrafast nano Fourier transform infrared (nano-FTIR) spectroscopy experimental system equipped with a 780 nm (~ 1.6 eV) pump laser and a tunable mid-infrared (MIR) probe laser, as shown in Fig. S1. Fig. 1a show the electric band structure of Ge[36], whose lowest point of valance band is at L valley with indirect band gap $E_{g, indirect}$ = 0.66 eV, and direct band gap $E_{g, direct}$ = 0.8 eV. Using the hv ≈ 1.6 eV pump photon energy, multi-valance band electric transition from HH band, LH band and SOH band will be excited, which marked by the red arrows in Fig. 1a. The influx of a large number of photo-induced electrons into the conduction band leads to the blue shift of the plasmon response of Ge nanowires, which increases the near-field signal intensity of the nanowire at higher frequency band. Fig. 1b show the five MIR probe bands (PBs) in our experiment, covering 700 – 1050 cm$^{-1}$ (PB-I), 1050 – 1400 cm$^{-1}$ (PB-II), 1300 – 1700 cm$^{-1}$ (PB-III), 1550 – 1980 cm$^{-1}$ (PB-IV) and 1650 – 2100 cm$^{-1}$

(PB-V). Fig. 1c shows the variation of the 3rd near-field amplitude s3(τ) in the PB-II with the pump-probe delay increases when the pump power is 20 mW. As shown in the inset of Fig. 1c, the 3rd near-field amplitude reaches maximum at τ = 1.8 ps, more than double the signal strength without photo-induced electrons injected (τ = -10 ps). Subsequently, the near-field signal falls back to the initial level within a time scale more than 20 ps. Around τ = 20 ps, the near-field signal intensity is slightly lower than the initial value, because the plasmon absorption of Ge nanowires falls back into the PB-II at this time. After that, the near-field signal intensity gradually returned to initial value due to electron-hole recombination. Compared with InAs and InSb[21,24,37], the peaking time of 1.8 ps and the relaxation time of 20 ps are relatively large, which is caused by the unique energy band structure of Ge. After the electrons in the valence band are excited to the conduction band, they first reach the X valley through intervalley electron scattering, and then reach the L valley[36], instead of directly reaching the L valley like InAs and InSb, which marked as the blue arrows in Fig. 1a.

We performed nano-FTIR to measure normalized amplitude spectra of Ge nanowire (Topography show in Fig. S2) s3/s3(Au) in Fig.1d with pump power increases. When there is no photogenerated electron injection (0 mW), the falling edge of near-field amplitude spectrum of the nanowires within the PB-I. As the pump power increases, this falling edge begins to blue-shift in a large range, and it is impossible to track the blue-shift process through a fixed PB. According to previous reports, the falling edge of the near-field amplitude spectrum in the Drude response of the free electron gas system corresponds to the plasmon absorption[38], so we tracked this falling edge by changing the PB to obtain the key information of the nanowire plasmons response turned by photo-induced electrons. As pump power increases from 0 mW to 10 mW, 20 mW and 30 mW, the falling edge blue shift from PB-I to PB-II, PB-III and PB-IV. When the pump power is greater than 30mW, the blue shift amplitude of the falling edge weakens, all within the PB-IV, which indicates that due to the limitation of the Pauli exclusion theorem, the saturation of the electronic states of the conduction band corresponding to the inter-band electronic transition. We collected transient spectra of the nanowire (Topography show in Fig. S2, not the same nanowire as above) plasmonic response at a pump power of 35mW in Fig. 1e. Likewise, we track the falling edge in the near-field amplitude spectrum by adjusting PBs. The pump power of 35mW was chosen instead of 50mW because too high pump power would cause thermal drift of the spatial position of the nanowires. When τ = -2 ps, the photo-induced electrons have not been injected yet, the falling edge of the spectrum is located within PB-II. When τ = 0 – 0.2 ps, it has a rapid and large-scale blue shift, and within PB-IV and PB-V, respectively. After that, the falling edge starts to redshift, redshifts to PB-IV at τ=4ps, redshifts to PB-III at τ=10ps, and returns to the initial position, PB-II at τ=30ps. In Fig. 1d-e, in order to avoid the influence of plasmon standing wave fringes that may be generated at the end of the nanowires on the spectral results, we collected spectra at a position away from the end of the nanowire. In addition, we use circles with the same color as each PB in Fig. 1b to represent the experimental data obtained in the corresponding PBs.

At the same time, the finite dipole model (FDM)[39] was used to calculate the nano-FTIR spectra to quantify the photo-induced plasmonic response. The permittivity of the Ge was presented as a Drude form[40]:

$$\varepsilon_{Ge} = \varepsilon_\infty \left(1 - \frac{\omega_p^2}{\omega^2 + i\gamma\omega}\right) \quad (1)$$

$\varepsilon_\infty = 16$ is the high-frequency permittivity[41], $\gamma$ is the damping rate (Here, we ignore the difference of $\gamma$ different energy bands.), $\omega_p = 2\pi\nu_p = \sqrt{\frac{ne^2}{m^*\varepsilon_0\varepsilon_\infty}}$ is the plasmon frequency, $n$ and $m^*$ represent the carrier density and effective mass, respectively. When there is no photo-induced electrons injection, all electrons in the conduction band of Ge nanowires are located in the L valley. At this time, the screened plasmon frequency can be expressed as: $\nu_p = \sqrt{\frac{n_{e,L}e^2}{m^*_{e,L}\varepsilon_0\varepsilon_\infty}}$, $n_{e,L}$ and $m^*_{e,L}$ represent the carrier density and effective mass contributed by L valley, respectively. After photo-induced electrons injection, the screened plasmon frequency can be written as: $\nu_p = \sqrt{\left(\frac{n_{e,L}}{m^*_{e,L}} + \sum_i \frac{n_{e,i}}{m^*_{e,i}} + \sum_i \frac{n_{h,i}}{m^*_{h,i}}\right)\frac{e^2}{\varepsilon_0\varepsilon_\infty}}$, i = HH, LH, SOH represent the contribution of HH band, LH band, and SOH band, $n_{e,i}$, $n_{h,i}$, $m^*_{e,i}$, $m^*_{h,i}$ represent the electrons and holes density and effective masses contributed by different valance band, respectively, $n_{e,i} =$

$n_{h,i}$. Since the effective masses of electrons and holes contributed by different energy bands are different, in the fitting process of the experimental spectrum using FDM method, in order to avoid overfitting due to too many fitting parameters, we set the plasmon frequency as the fitting parameters instead of using the carrier density contributed by each energy band as a fitting parameter. The black solid lines in Fig. 1d-e are the fitting results and the extracted plasmon frequencies are shown in Fig. 1f–g. In Fig. 1f, as the pump power increased from 0 mW to 30 mW, the $v_p$ blue shifted rapidly from 860 cm$^{-1}$ to 1620 cm$^{-1}$. When the pump power reached 30 mW, hot electrons occupied most states in the Γ-valley, and the pump absorption efficiency decreased intensely. When the power contacts 50 mW, the $v_p$ reaches the saturation value of 1700 cm$^{-1}$. In Fig. 1g, at τ = 0.2 ps, $v_p$ = 2048 cm$^{-1}$ reaches the maximum value. As the delay τ increases, $v_p$ decreases rapidly, and at τ = 12 ps, $v_p$ =1350 cm$^{-1}$, which decrease of about 34%. For InSb, the time it takes for $v_p$ to decrease by the same magnitude is 4 ps[24]. Again, this is due to the intervalley electron scattering (Γ→X→L) in the Ge conduction band mentioned above[36]. The slower falling speed helps us to control its plasmon more calmly, which has great significance for the application of ultrafast nanophotonic devices.

In Fig. 1d, the plasmon frequency of Ge nanowire $v_p$ = 860 cm$^{-1}$ when pump power is 0 mW, and in Fig. 1e, $v_p$ = 1134 cm$^{-1}$ when τ = -2 ps, both of these conditions have no photo-induced electrons injected. The reason for this difference is the difference in electron density among individual Ge nanowires prepared by chemical vapor deposition (CVD) method. The intrinsic electron density of the nanowire can be estimated from the plasmon frequency. The effective mass of electrons in L valley of Ge is $m_L^*$ = 0.12$m_0$[36], $m_0$ is electron mass. For $v_p$ = 860 cm$^{-1}$, $n_{e,L}$ = 1.58 × 10$^{19}$ cm$^{-3}$, and for $v_p$ = 1134 cm$^{-1}$, $n_{e,L}$ = 2.75 × 10$^{19}$ cm$^{-3}$. Due to the large indirect band gap of Ge, the multi-valence band electronic transitions excited by the 1.6 eV excitation energy all occur near the Γ point, which can be approximately regarded as occurring in the spherical energy band near the Γ point of the conduction band. Based on this, we can estimate that the injection of photo-induced carriers increases the carrier density of Ge nanowire by about $4 \times 10^{19}$ cm$^{-3}$(See section 3 in SI)[42–44].

The ultra-broadband tuning range of Ge nanowires plasmon makes it possible to support propagating SPPs. We conducted near-field imaging with the MIR probe PB-III to visualize the standing wave pattern of the plasmonic F-P resonance on a Ge nanowire with 90 nm diameter is shown in Fig. 2a. PB-III is far away from the phonon resonance of SiO$_2$, which can avoid the influence of SiO$_2$ phonons on SPPs. Figure 2b shows near-field optical images with increasing pump power. As the pump power increases, the injection of photo-induced carriers shifts the frequency range of SPP into PB-III. When the pump power exceeds 30mW, the standing wave stripes of SPPs gradually appear in the nanowire, and the brightness of the standing wave stripes reaches the strongest when the pump power is 50mW which is 4.5 times of the 3rd near-field amplitude of SiO$_2$. The distribution of light and dark nods in the near-field images of 30 − 50 mW pump power is different from the SPPs standing wave patterns in the single-frequency near-field images[32,45]. It is the envelope formed by the superposition of SPP fringes at different frequencies. We extract the line profiles along the long axis of the nanowire (red dashed line in Fig. 2b), as shown in Fig. 2c. There is no envelope at 0 mW, and 4 obvious envelopes at 50 mW. Fig. 2d shows near-field optical images of nanowires within the PB-III as the pump-probe delay τ increases. At τ = 0 ps, the near-field signal of the nanowires increases suddenly, and the envelope of standing wave fringe appears. With the saturation of the photo-induced carriers, the near-field response of the Ge nanowires reaches the strongest at τ = 0.2 ps. The Ge nanowires maintain a high near-field response for about 5 ps due to intervalley electron scattering, and then, the near-field response gradually decreases due to carrier recombination.

The near-field images, as presented in Fig. 2b and 2d, cannot account for the parameter of SPPs at a given frequency. We conducted frequency-resolved hyperspectral line scans to isolate the frequency-dependent SPPs embedded within the standing wave pattern. For this study, we selected a Ge nanowire with a diameter of 75 nm (Topography show in Fig. S4). Hyperspectral images of the different pump power are presented in Fig.3a-b. There is no standing wave fringe in the result of 0 mW, but at 35 mW, three obvious standing wave fringes appear in the detection range, as indicated by the black dashed lines in Fig. 3b. In Fig. 3b, as the frequency increases, the distance between three standing wave fringes gradually decreases, and the plasmon wavelength decreases. We use the finite element software COMSOL 5.2a to analyze the mode of the nanowire plasmons, and find that the SPPs observed in the experiment are $TM_0$ mode, as show in Fig. 3c.

**Conclusion**

In conclusion, we have demonstrated the activation and manipulation of Ge nanowire plasmons through optical excitation. Multi-valence band electron transitions result in ultra-broadband plasmon tuning across 800 – 2000 $cm^{-1}$. Using optical tunable method, we have realized the plasmonic response of semiconductors above 1500 $cm^{-1}$ for the first time. Additionally, combining near-field broadband imaging and hyperspectral line scan, we have achieved the imaging of the propagating SPPs of semiconductor nanowire tuned by photo-induced carriers for the first time. On the one hand, the mature preparation process of Ge makes it widely used in semiconductor optoelectronic devices. On the other hand, the biocompatibility of Ge nanowires is good, and the plasmonic response in the band higher than 1500$cm^{-1}$ makes it have great application prospects in the field of protein and nucleic acid detection. These findings open new avenues for developing novel optoelectronic devices based on semiconductor plasmons.

Figures

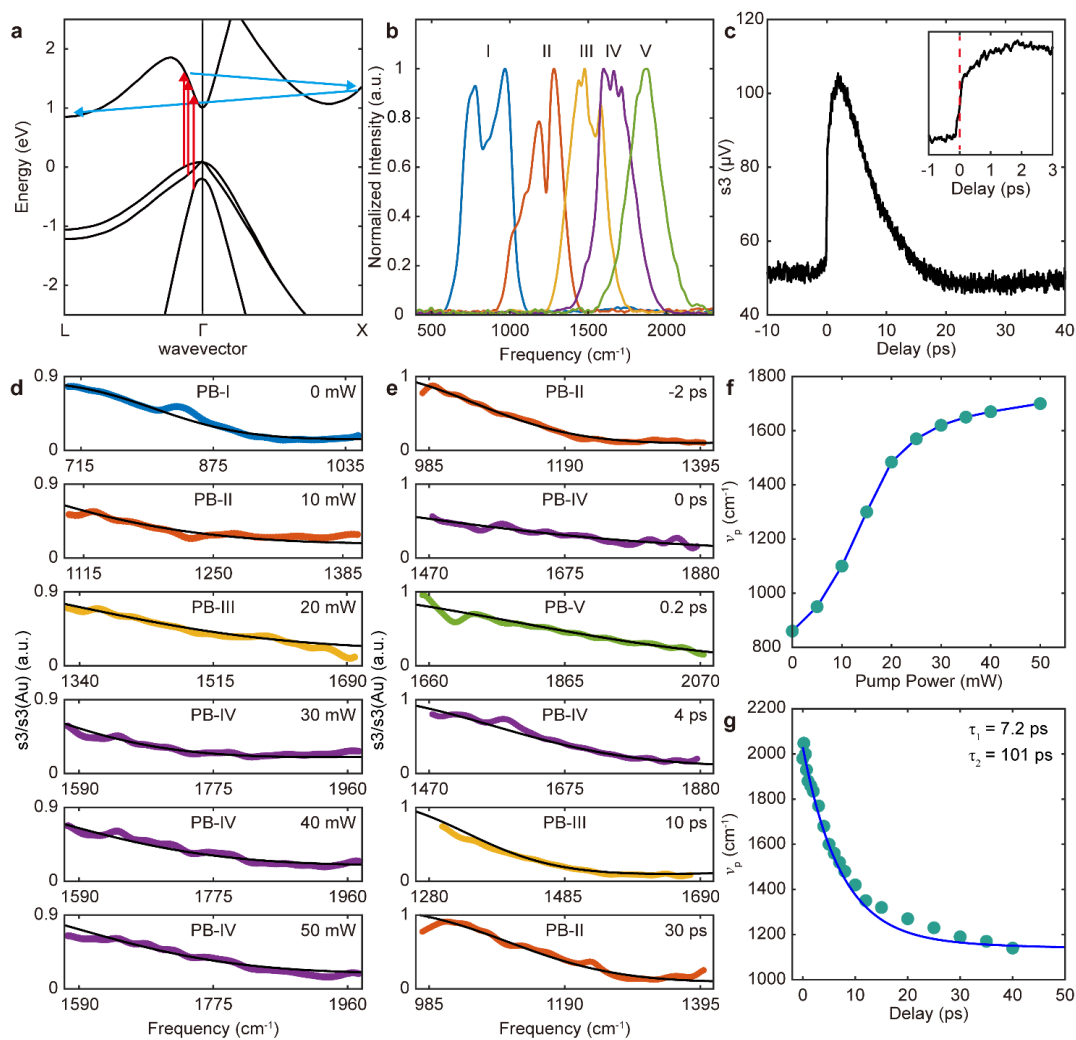

**Fig. 1 Ultrafast dynamics and optically tunable plasmon response of individual Ge nanowire.** (a) The electronic band structure and the pump induced electronic transition. Red arrows are the electronic transitions from the HH, LH, and SOH, respectively. Blue arrows are intervalley scattering of Γ to X and X to L. (b) Near field amplitude spectra of five PBs. (c) Pump-probe amplitude scans on the Ge nanowire in PB-II. (d) Near-field amplitude spectrum s3/s3(Au) for different pump power. From top to bottom, the pump power increases from 0 to 50 mW. (e) Transient near-field amplitude spectrum s3/s3(Au) for the pump power of 35 mW. From top to bottom, the pump-probe delay increases. Opened diamonds are the experimental data, and solid lines are fitting results using the FDM. Traces are vertically offset for clarity. In d-e, the circles in different colors are the experimental data within the PBs corresponding to b, and the solid black lines are fitting results using the FDM. Traces are vertically offset for clarity. (f) The photo-induced plasmon frequency $\nu_P$ for a pump power of 0 – 50 mW. (g) The relaxation

of the plasmon frequency $v_p$ for a delay time of $\tau = 0 - 40$ ps. The plasmon frequency $v_p$ is presented with biexponential fits.

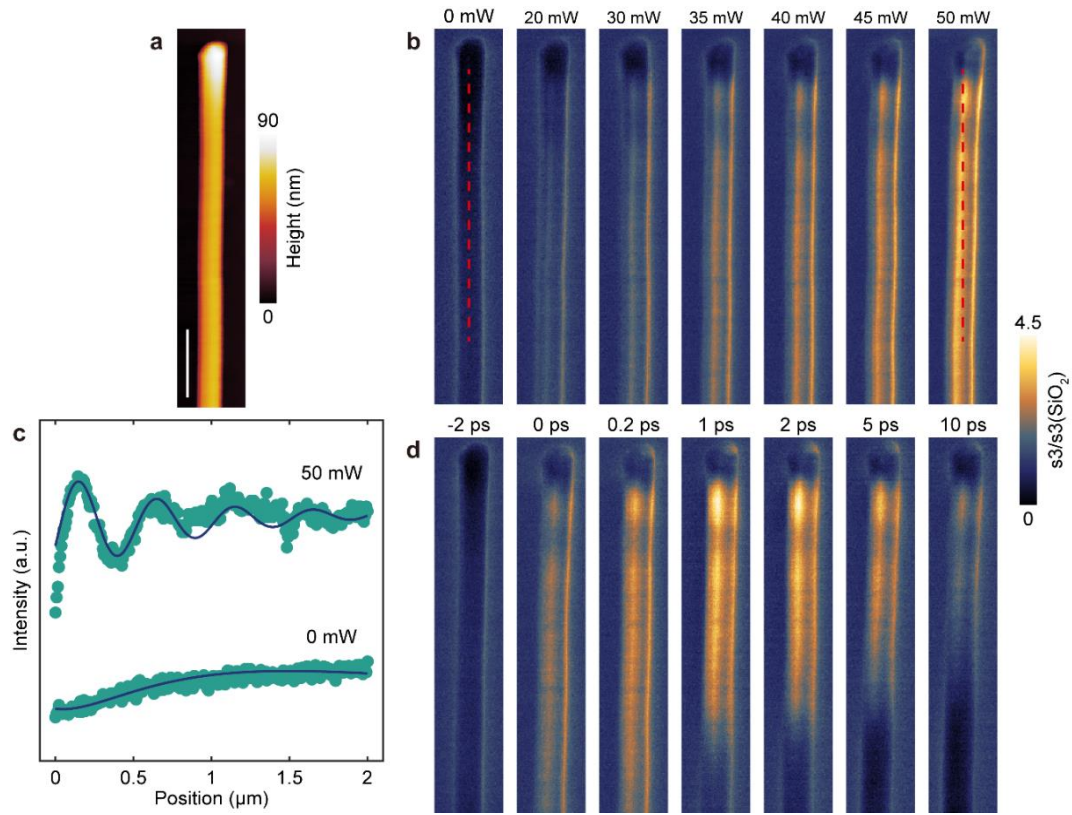

**Fig. 2 Broadband near-field images of the Ge nanowire SPPs.** (a) Topography image of the Ge nanowire. (b) Spatial maps of the third harmonic scattered near-field amplitude s3/s3(SiO$_2$) for the pump power of 0 – 50 mW. (c) Line profiles of Ge nanowire near-field images under 0mW and 50mW pump power from b (red dashed line). (d) Spatial maps of s3/s3(SiO$_2$) for $\tau$ = -2 – 10 ps. The pump power is 35 mW. The near-field images in b and d are both obtained in PB-III.

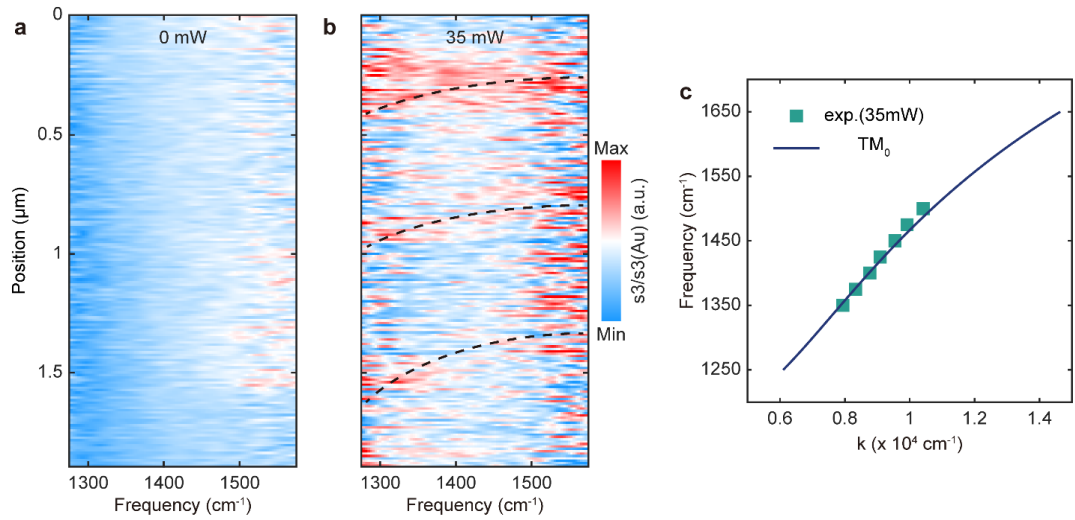

**Fig. 3 Hyperspectrum images of the Ge nanowire SPPs**. (a-b) Hyperspectral line scans along the long axis of nanowire for the pump power of 0 mW and 35 mW. Both of these results are obtained in PB-III. (c) Plasmon dispersion of Ge nanowire. Green squares are experiment data extracted from b and black line is the dispersion of $TM_0$ mode of Ge nanowire plasmons.